\documentclass{jetpl}
\twocolumn

  \title{Evolution of cosmological constant in effective
gravity}

\rtitle{Evolution of cosmological constant in effective
gravity}

\sodtitle{Evolution of cosmological constant in effective
gravity}

\author{G.\,E.\,Volovik\/\thanks{e-mail:
volovik@boojum.hut.fi}}

\rauthor{Volovik G.E.}

\sodauthor{Volovik}

\address{Low Temperature Laboratory, Helsinki University of 
Technology, P.O.Box 2200, FIN-02015, HUT, Finland\\~\\
Landau Institute for Theoretical Physics RAS, Kosygina 2, 117334 
Moscow, Russia}

\dates{20 February 2003}{*}

\abstract{In contrast  to the phenomenon of nullification  of
the cosmological constant in the equilibrium vacuum,
which is the general property of any quantum vacuum,
there are many options in modifying the Einstein
equation to allow the  cosmological constant to
evolve in a non-equilibrium vacuum. An attempt is made
to extend the Einstein equation in the direction
suggested by the condensed-matter analogy of the
quantum vacuum. Different scenarios are found depending
on the behavior of and the relation between the
relaxation parameters involved, some of these scenarios
having been discussed in the literature. One of them
reproduces the scenario in which the effective
cosmological constant emerges as a constant of
integration. The second one describes the situation,
when after the cosmological phase transition the
cosmological constant drops from zero to the negative
value; this scenario describes the relaxation from this
big negative value back to zero and then to a small
positive value. In the third example the relaxation
time is not a constant but depends on matter; this
scenario demonstrates that the vacuum energy (or its
fraction) can play the role of the cold dark matter.}

\begin{document}

\maketitle

  {\bf 1. Introduction.}

It is clear that the  pressure of the vacuum in our
universe is very close to zero as
compared to the Planck energy scale, and thus the
experimental cosmological constant is close to zero.
However,  if at this almost zero pressure one starts
calculating the vacuum energy by summing all the
positive and negative energy states, one obtains a huge
vacuum energy which by about 120 orders of
magnitude exceeds the experimental limit. This is the
main cosmological constant problem
\cite{Weinberg2,Sahni,Peebles,Padmanabhan}.

Exactly the same `paradox' occurs in any quantum liquid
(or in any other condensed matter) at zero pressure. The
experimental energy of the ground state of, say, the
quantum liquid at zero pressure is zero. On the other
hand, if one starts calculating the vacuum energy
summing the energies of all the positive and negative
energy modes up to the corresponding Planck (Debye)
scale, one obtains the huge energy. However, there is
no real paradox in quantum liquids, since if one adds
all the trans-Planckian (microscopic, atomic) modes one
immediately obtains the zero value, irrespective of the
details of the microscopic physics \cite{VolovikBook}.
The fully microscopic consideration restores the
Gibbs-Duhem relation $\epsilon=-p$ between the  energy
(the relevant thermodynamic potential) and the pressure
of the quantum liquid at $T=0$, which ensures that the
energy of the vacuum state $\epsilon=0$ if the external
pressure is zero.

This is the first message from condensed matter to the
physics of the quantum vacuum:  One should not worry
about the huge vacuum energy, the trans-Planckian
physics with its degrees of freedom will do all the
job of the cancellation of the vacuum energy without
any fine tuning and irrespective of the details of the
trans-Planckian physics.   There are
other messages which are also rather
general and do not depend much on details of the
trans-Planckian physics.  For example,
if the cosmological constant is zero above the
cosmological phase transition, it will become zero
below the transition after some transient period.

Thus from the quantum-liquid analog of the quantum
vacuum it follows that the  cosmological constant is
not a constant but is an evolving physical parameter,
and our goal is to find the laws of its evolution.
In contrast to the phenomenon of the cancellation  of
the cosmological constant in the equilibrium vacuum,
which is the general property of any quantum vacuum,
there are many options in modifying the Einstein
equation to allow the  cosmological constant to
evolve. However,  the condensed
matter physics teaches us that we must avoid the
discussion of the microscopic models of the quantum
vacuum \cite{Laughlin} and use instead the general
phenomenological approach. That is why we do not
follow the traditional way of description in
terms of, say, the scalar field which mediates
the decay of the dark energy \cite{Alam}, and present an
attempt of the  pure
phenomenological description by introducing the
dissipative terms directly into the Einstein equation.

{\bf 2. Einstein equation.}

  {\it Standard formulation}.--
Let us start with the non-dissipative equation for
gravity -- the Einstein equation. It is
obtained from the action:
\begin{equation}
S=S_{\rm E}+S_\Lambda+S_{\rm M}~~'
\label{EinsteinAction1}
\end{equation}
where $S_{\rm M}$ is the matter action;
\begin{equation}
S_{\rm E}= -\frac{1}{16\pi G}\int d^4x\sqrt{-g}{\cal
R}~~'
\label{EinsteinAction2}
\end{equation}
is the Einstein curvature action; and
\begin{equation}
  S_\Lambda=
-\frac{\Lambda}{ 8\pi G}\int d^4x\sqrt{-g}~,
\label{EinsteinAction3}
\end{equation}
where
$\Lambda$  is the cosmological constant
\cite{EinsteinCosmCon}. Variation over the metric
$g^{\mu\nu}$ gives
\begin{equation}
  \frac{1}{ 8\pi G}\left(G_{\mu\nu}  -
\Lambda g_{\mu\nu}\right)
  =T^{\rm M}_{\mu\nu}~,
\label{EinsteinEquation}
\end{equation}
where
\begin{equation}
  G_{\mu\nu}=
R_{\mu\nu}- \frac{1}{2}{\cal R}g_{\mu\nu}  ~
\label{EinsteinTensor}
\end{equation}
is the Einstein tensor, and
$T^{\rm M}_{\mu\nu}$ is the
energy-momentum tensor for matter. This form of the
Einstein equation implies that the matter fields on
the right-hand side of Eq.(\ref{EinsteinEquation}) serve
as the source of the gravitational field, while the
$\Lambda$-term belongs to the gravity.

Moving the  $\Lambda$-term to the
rhs of the Einstein equation changes the meaning of the
cosmological constant.  The $\Lambda$-term becomes
the energy-momentum tensor of the vacuum, which in
addition to the matter is the source of the
gravitational field
\cite{Bronstein}:
\begin{equation}
   \frac{1}{ 8\pi
G}G_{\mu\nu}=T^{\rm
M}_{\mu\nu}+T^\Lambda_{\mu\nu}~,~ T^\Lambda_{\mu\nu}
=\rho_\Lambda g_{\mu\nu} = \frac{\Lambda}{
8\pi G}g_{\mu\nu}~.
\label{EinsteinEquation10}
\end{equation}
Here $\rho_\Lambda$ is
the vacuum energy density and $p_\Lambda=-\rho_\Lambda$
is the vacuum pressure.

{\it Einstein equation in induced gravity}.--
In the induced gravity introduced by
Sakharov \cite{Sakharov}, the gravity is the
elasticity of the vacuum, say, fermionic vacuum, and
the action for the gravitational field is induced by
the vacuum fluctuations of the fermionic matter fields.
Such kind of the effective gravity emerges in quantum
liquids
\cite{VolovikBook}. In the induced gravity  the Einstein
tensor must be also moved to the matter side, i.e. to
the rhs:
\begin{equation}
0=T^{\rm
M}_{\mu\nu}+T^\Lambda_{\mu\nu}+T^{\rm curv}_{\mu\nu}~.
\label{EinsteinEquation2}
\end{equation}
where
\begin{equation}
  T^{\rm curv}_{\mu\nu}
=- \frac{1}{ 8\pi G}G_{\mu\nu}
  ~
\label{Curvature}
\end{equation}
has the meaning of the energy-momentum tensor produced
by  deformations of the fermionic vacuum. It
describes such elastic deformations of the vacuum, which
distort the effective metric field $g_{\mu\nu}$ and
thus play the role of the gravitational field. As
distinct from the
$T^\Lambda_{\mu\nu}$ term which is of the 0-th order in
gradients of the metric, the $T^{\rm curv}_{\mu\nu}$
term is of the 2-nd order in gradients of
$g_{\mu\nu}$. The  higher-order gradient terms
also naturally appear in induced gravity.

In the induced
gravity the free gravitational field is absent, since
there is no gravity in the absence of the quantum
vacuum. Thus the total energy-momentum tensor comes
from the original (bare) fermionic degrees of freedom.
That is why all the the contributions to the
energy-momentum tensor are obtained by the variation of
the total fermionic action over
$g^{\mu\nu}$:
\begin{equation}
  T^{\rm total}_{\mu\nu}= \frac{2}{  \sqrt{-g}}
\frac{\delta S}{\delta g^{\mu\nu}}= - \frac{1}{ 8\pi
G}G_{\mu\nu}  +
\rho_\Lambda g_{\mu\nu} +T^{\rm M}_{\mu\nu} ~.
\label{TotalEMT}
\end{equation}
According to the variational principle, $\delta
S/\delta g^{\mu\nu}=0$, the total
energy-momen\-tum tensor is zero, which gives rise to
the Einstein equation in the
form of Eq.(\ref{EinsteinEquation2}). Since $T^{\rm
total}_{\mu\nu}=0$, it
satisfies the conventional and covariant conservation
laws,  $T^{{\rm
total}~\nu}_{\mu,\nu}=0$ and $T^{{\rm
total}~\nu}_{\mu;\nu}=0$, and thus serves as the
covariant and localized energy-momentum tensor of
matter and gravity.

In this respect there is no much difference between
different contributions to the energy-momentum tensor:
all come from the original fermions. However, in the
low-energy corner, where the gradient expansion
for the effective action works, one can distinguish
between different contributions: (i) some part of the
energy-momentum tensor ($T^{\rm M}_{\mu\nu}$) comes
from the excited fermions -- quasiparticles -- which in
the effective theory form the matter. The other parts
come from the fermions forming the vacuum -- the Dirac
sea. The contribution from the vacuum fermions
contains: (ii) The zeroth-order term in the gradients
of $g_{\mu\nu}$; this is the energy-momentum tensor of
the homogeneous vacuum -- the
$\Lambda$-term. Of course, the whole Dirac sea cannot
be sensitive to the change of the effective
infrared fields $g_{\mu\nu}$: only small
infrared perturbations of the vacuum, which we are
interested in, are described by these effective fields.
(iii)  The stress tensor of the inhomogeneous
distortion of the vacuum state, which plays the role of
gravity; the second-order term
$T^{\rm curv}_{\mu\nu}$ in the stress tensor represents
the curvature term in the Einstein equation.

The same occurs in induced QED \cite{Zeldovich1967b},
where the electromagnetic field is induced by the vacuum
fluctuations of the same fermionic field. The total
electric current
\begin{equation}
j^\mu= \frac{\delta
S}{\delta A_\mu}=  \frac{\delta
S^{\rm int}}{\delta A_\mu}+ \frac{\delta
S^{\rm Maxwell}}{\delta A_\mu}= j^\mu_{\rm
charged~particles} +  j^\mu_{\rm field},
\label{TotalCurrent}
\end{equation}
is produced by excited fermions (the 1-st term) and by
fermions in the quantum vacuum  (the 2-nd term). The
electric current in the second term is produced by such
elastic deformations of the fermionic vacuum which play
the role of the electromagnetic field, and the
lowest-order term in the effective action describing
such distortion of the vacuum state is the induced
Maxwell action
\begin{equation}
S^{\rm Maxwell}=
\int dt d^3x  \frac{\sqrt{-g}}{ 16\pi
\alpha}F^{\mu\nu}F_{\mu\nu}~,
\label{MaxwellAction}
\end{equation}
which is of the second order in gradients of
$A_\mu$.   According to the
variational principle, $ \delta
S/\delta A_\mu=0$, the total electric current
produced by the vacuum and excited fermions is zero.
This means that the system is always locally
electro-neutral. This ensures that the
homogeneous vacuum state  without
excitations has zero electric charge, i.e. our
quantum vacuum is electrically neutral. The same occurs
with the hypercharge, weak charge, and color charge of
the vacuum: they are zero in the absence of matter and
fields.

In
the traditional approach the cosmological constant is
fixed, and it serves as the source for the metric
field: in other words the input in the Einstein equation
is the cosmological constant, the output is  de Sitter
expansion, if matter is absent.
In effective gravity,
where the gravitational field, the matter fields, and
the cosmological `constant' emerge simultaneously in the
low-energy corner, one cannot say that one of these
fields is primary and serves as a source for the
other fields thus governing their behavor. The
cosmological constant, as one of the players,  adjusts
to the evolving matter and gravity in a self-regulating
way. In particular, in the absence of matter ($T^{\rm
M}_{\mu\nu}=0$) the non-distorted vacuum ($T^{\rm
curv}_{\mu\nu}=0$) acquires zero cosmological constant,
since, according to the `gravi-neutrality' condition
Eq.(\ref{EinsteinEquation2}), it follows from equations
$T^{\rm M}_{\mu\nu}=0$ and $T^{\rm curv}_{\mu\nu}=0$
that
$T^{\Lambda}_{\mu\nu}=0$. In this approach, the
input is the vacuum configuration (in a given example
there is no matter, and the vacuum is homogeneous), the
output is the vacuum energy. In contrast
to the traditional approach, here the gravitational
field and matter serve as a source of the induced
cosmological constant.

This conclusion is supported by the effective gravity
and effective QED which emerge in quantum liquids
or any other condensed matter system
of the special universality class \cite{VolovikBook}.
The nullification of the vacuum energy in the
equilibrium homogeneous vacuum state of the system also
follows from the variational principle, or more
generally from the Gibbs-Duhem relation applied to the
equilibrium vacuum state of the fermionic system if it
is isolated from the environment. In the absence of
the environment one has $p_\Lambda=0$, while from
the Gibbs-Duhem relation $\rho_\Lambda=-p_\Lambda$ at
$T=0$ it follows that
$\rho_\Lambda=0$. This corresponds to
$T^{\Lambda}_{\mu\nu}=0$ for quiescent flat Universe at
$T=0$, i.e. quiescent flat Universe without matter is
not gravitating.

  {\bf 3. Modification of Einstein equation  and
relaxation of the vacuum energy.}

{\it Dissipation in Einstein equation}.--The Einstein
equation does not allow us to obtain the time
dependence of the cosmological constant, because of the
Bianchi identities $G^{\nu}_{\mu;\nu}=0$ and covariant
conservation law for matter fields (quasiparticles)
$T^{\nu{\rm M}}_{\mu;\nu}=0$ which together lead to
$\partial_\mu\Lambda=0$. But they allow us to obtain
the value of the cosmological constant in different
static universes, such as the Einstein closed  Universe
\cite{EinsteinCosmCon}, where the cosmological constant
is obtained as a function of the curvature and matter
density. To describe the evolution of the cosmological
constant the relaxation term must be added.

The dissipative
term in the Einstein equation  can be introduced in
the same way as  in two-fluid
hydrodynamics \cite{Khalatnikov} which serves as the
non-relativistic analog of the self-consistent
treatment of the dynamics of vacuum (the superfluid
component of the liquid) and matter (the normal
component of the liquid) \cite{VolovikBook}
\begin{equation}
  T^{\rm
M}_{\mu\nu}+T^\Lambda_{\mu\nu}+T^{\rm curv}_{\mu\nu}+
T^{\rm diss}_{\mu\nu}=0~,
\label{EinsteinEquationDissipation}
\end{equation}
where $T^{\rm diss}_{\mu\nu}$ is the dissipative part
of the total energy-momentum
tensor. In contrast to the conventional dissipation of
the matter, such as viscosity and thermal conductivity
of the cosmic fluid, this term is not the part of
$T^{\rm M}_{\mu\nu}$. It describes the dissipative
back reaction of the vacuum, which does not influence
the matter conservation law $T^{\nu{\rm
M}}_{\mu;\nu}=0$. The condensed-matter example of such
relaxation of the variables describing the fermionic
vacuum is provided by the dynamic equation for the
order parameter in superconductors -- the
time-dependent Ginzburg-Landau equation  which contains
the relaxation term (see e.g. the book
\cite{KopninBook}).

Let us
consider how the relaxation occurs on the example of the
  spatially flat Robertson-Walker metric
\begin{equation}
ds^2=dt^2-a^2(t) d{\bf r}^2~.
\label{RobertsonWalkerFlat}
\end{equation}
The Ricci tensor and scalar and
the Einstein tensor are
\begin{equation}
R_0^0= -3 \frac{\partial_t^2a}{ a} ~~;~~ R_j^i=
-\delta^i_j
\left( \frac{\partial_t^2a}{ a} + 2 \frac{(\partial_t
a)^2}{ a^2}\right)~,
\label{CurvatureTensor}
\end{equation}
\begin{equation}
{\cal R}= -6 \left( \frac{\partial_t^2a}{
a} +  \frac{(\partial_t a)^2}{ a^2}\right)~,
\label{RicciScalar}
\end{equation}
\begin{equation}
  G_0^0=
R_0^0- \frac{1}{ 2}{\cal R} =3 \frac{(\partial_t
a)^2}{ a^2}\equiv 3H^2~, ~
\label{EinsteinTensorRW00}
\end{equation}
\begin{equation}
  G_i^j=
R_i^j- \frac{1}{ 2}{\cal
R}\delta_i^j=\delta_i^j\left(H^2 +2  \frac{ \partial_t^2
a }{ a}\right)=\delta_i^j\left(3H^2 +2\dot H\right).
\label{EinsteinTensorRWij}
\end{equation}

In the lowest order of the gradient expansion, the
dissipative part $T^{\rm diss}_{\mu\nu}$ of the stress
tensor describing the relaxation  of
$\Lambda$ must be proportional to the first time
derivative.

{\it Cosmological constant as integration
constant}.--Let us start with the following guess
\begin{equation}
T^{\rm diss}_{\mu\nu}= \tau_\Lambda
  \frac{\partial  \rho_\Lambda}{\partial t}
g_{\mu\nu}~.
\label{DissipationTensorLambda}
\end{equation}
This non-covariant term implies the existence
of the preferred reference frame, which is the natural
ingredient of the trans-Planckian physics where the
general covariance is violated.
The modified Einstein equation in the absence of
matter are correspondingly
\begin{equation}
  3H^2=\Lambda +\tau_\Lambda \dot \Lambda~, ~
\label{EinsteinTensorRW00}
\end{equation}
\begin{equation}
  3H^2+2\dot H=\Lambda +\tau_\Lambda \dot \Lambda~.
\label{EinsteinTensorRW00}
\end{equation}
This gives the constant Hubble parameter $H=$ constant,
i.e. the exponential de Sitter expansion or
contraction. The cosmological `constant'
$\Lambda$ relaxes to the value determined by the
expansion rate:
\begin{equation}
  \Lambda(t)=3H^2 +\left( \Lambda(0)
-3H^2\right)\exp\left(- \frac{t}{
\tau_\Lambda}\right)~.
\label{EvolutionLambda}
\end{equation}
This is consistent with the Bianchi identity, which
requires that $\partial_t(\Lambda +\tau_\Lambda \dot
\Lambda)=0$. Actually this situation corresponds to the
well known case when the cosmological constant arises
as an integration constant (see reviews
\cite{Weinberg2,Padmanabhan}). Here it is the
integration constant
$\Lambda_0= \Lambda +\tau_\Lambda
\partial_t\Lambda$. Such scenario emerges because
the dissipative term in
Eq.(\ref{DissipationTensorLambda}) is proportional to
$g_{\mu\nu}$.

{\it Model with two relaxation parameters}.--
Since $T^{\rm diss}_{\mu\nu}$ is tensor, the
general description of the vacuum relaxation requires
introduction of several relaxation times. This also
violates the Lorentz invariance, but we already assumed
that the dissipation of the vacuum variables due to
trans-Planckian physics implies the existence of the
preferred reference frame. In the isotropic space we
have only two relaxation times: in the energy and
pressure sectors. In the presence of
matter one has
\begin{equation}
  3H^2=\Lambda +\tau_1 \dot \Lambda + 8\pi
G\rho^{\rm M}~, ~
\label{EinsteinTensorRW00M}
\end{equation}
\begin{equation}
  3H^2+2\dot H=\Lambda +\tau_2 \dot \Lambda- 8\pi
Gp^{\rm M}~.
\label{EinsteinTensorRWijM}
\end{equation}
  Since the covariant
conservation law for matter does not follow now from the
Bianchi identities, these two equations
must be supplemented  by the covariant conservation law
to prevent the creation of matter:
\begin{equation}
a \frac{\partial}{ \partial a}(\rho^{\rm
M}a^3)=p^{\rm M}a^3~.
\label{CovariantConservation}
\end{equation}

Let us consider the simplest case when the
relaxation occurs only in the pressure
sector, i.e. $\tau_1=0$. We assume also that the
ordinary matter is cold, i.e. its pressure $p^{\rm
M}=0$, which gives
$\rho^{\rm M}\propto a^{-3}$. Then one finds two
classes of solutions: (i)
$\Lambda={\rm constant}$; and (ii)
$H=1/(3\tau_2)$. The first one corresponds to the
conventional expansion with constant $\Lambda$-term and
cold matter, so let us discuss the second solution,
$H=1/(3\tau_2)$.

{\it Relaxation after cosmological phase
transition}.--
In the simplest case
when
$\tau_2$ is constant, $\tau_2\equiv
\tau={\rm constant}$, the $\Lambda$-term and the energy
density of matter
$\rho^{\rm M}$ exponentially relax to $1/(3\tau^2)$ and
to 0 respectively:
\begin{equation}
H= \frac{1}{ 3\tau},~\Lambda(t)= \frac{1}{ 3\tau^2}
-  8\pi G\rho^{\rm M}(t), ~  \frac{\rho^{\rm M}(t)}{
\rho^{\rm M}(0)}= \exp\left(- \frac{t}{ \tau}\right).
\label{Evolution}
\end{equation}
Such solution describes the behavior after the
cosmological phase transition. According to the
condensed-matter example of the phase transition,
the cosmological `constant' is (almost) zero before the
transition; while after the transition it drops to the
negative value, and then relaxes back to zero
\cite{VolovikBook}. Eq.(\ref{Evolution}) corresponds to
the latter stage, but it demostrates that in its
relaxation after the phase transition the
$\Lambda$-term  crosses zero and finally becomes
a small positive constant determined by
the relaxation parameter $\tau$ which governs the
exponential de Sitter expansion.

{\it Dark energy as dark matter}.--
Let us now allow  $\tau$ to vary. Usually the
relaxation and dissipation are determined by matter
(quasiparticles).  The term which violates
the Lorentz symmetry or the general covariance
must contain the Planck scale $E_{\rm Planck}$ in the
denominator, since it must disappear at infinite Planck
energy. The lowest-order term,   which contains the
$E_{\rm Planck}$ in the denominator, is
$\hbar/\tau\sim
T^2/ E_{\rm Planck}$, where $T$ is the characteristic
temperature or energy of matter. In case of radiation it
can be written in terms of the radiation density:
\begin{equation}
   \frac{1}{  3\tau^2}=8\pi\alpha G\rho^{\rm M}~,
\label{MatterDependentTau}
\end{equation}
where $\alpha$ is the dimensionless parameter. If
Eq.(\ref{MatterDependentTau}) can be applied to the
cold baryonic matter too, then the solution of the
class (ii) becomes again
$H=1/(3 \tau)$, but now $\tau$ depends on the
matter field.  This solution gives the standard power
law for the expansion of the cold flat universe and the
relation between
$\Lambda$ and the baryonic matter $\rho^{\rm M}$:
\begin{equation}
a \propto           t^{2/3} ,~8\pi G \rho^{\rm
M}=  \frac{4}{
3 \alpha  t^2},~H= \frac{2}{
3t},~\Lambda=(\alpha-1)8\pi\ G\rho^{\rm M}.
\label{DarkEnergyMatterRelation}
\end{equation}
In terms of the densities
normalized to $\rho_c=3H^2/8\pi G$ (the
critical density corresponding to the flat universe in
the absence of the vacuum energy)
$\Omega_\Lambda=
\rho_\Lambda/\rho_c$ and $\Omega^{\rm M}=\rho^{\rm
M}/\rho_c$ one has
\begin{equation}
\Omega^{\rm M}= \frac{1}{
\alpha}~,~\Omega_\Lambda= \frac{\alpha-1}{
\alpha}~.
\label{NormalizedDarkEnergyMatterRelation}
\end{equation}
Since  the effective vacuum pressure in
Eq.(\ref{EinsteinTensorRWijM}) is $p_\Lambda\propto
-(\Lambda+\tau\dot\Lambda)=0$, in this
solution the dark energy  behaves as the cold dark
matter. Thus the vacuum energy can serve as  the origin
of the non-baryonic dark matter.

{\bf 4. Conclusion.}

In the effective gravity the equilibrium
time-independent vacuum state without matter is
non-gravitating, i.e. its relevant vacuum energy which
is responsible for gravity is zero. In a non-equilibrium
situation the cosmological constant is non-zero, but it
is an evolving parameter rather than the constant. The
process of relaxation of the cosmological constant,
when the vacuum is disturbed and out of the
equilibrium, requires some modification of the Einstein
equation violating the Bianchi identities to allow the
cosmological constant to vary.   In contrast to the
phenomenon of nullification  of the cosmological
constant in the equilibrium vacuum, which is the general
property of any quantum vacuum and does not depend on
its structure and on details of the trans-Planckian
physics, the deviations from the general relativity can
occur in many different ways, since  there are many
routes from the low-energy effective theory to the
high-energy `microscopic' theory of the quantum vacuum.
However, it seems reasonable that such modification can
be written in the general phenomenological way, as for
example the dissipative terms are introduced in the
hydrodynamic theory. Here we suggested to describe the
evolution of the $\Lambda$-term by two
phenomenological parameters (or functions) -- the
relaxation times. The corresponding dissipative terms in
the stress tensor of the quantum vacuum are determined
by trans-Planckian physics and do not obey the
general covariance.

We discussed here simplest examples of the
relaxation of the vacuum to equilibrium described by
a single relaxation parameter. The
first example ($\tau_1=\tau_2$) reproduces the
well known scenario in which the effective cosmological
constant emerges as a constant of integration. The
second example ($\tau_1=0$ and
$\tau_2=$ constant) describes the situation which
occurs if, after the cosmological phase transition,
$\Lambda$ acquires a big negative value:
$\Lambda$ relaxes back to zero and then to a small
positive value. The third example, when
$\tau_2$ is determined by the baryoinic matter density,
demonstrates that the vacuum energy (or its fraction)
can play the role of the cold dark matter.

These
examples are too simple to describe the real
evolution of the present universe and are
actually excluded by observations \cite{Sahni}. The
general consideration with two  relaxation functions
  is needed. In this general case, it
corresponds to the varying in time parameter
$w_Q=p_Q/\rho_Q$ describing the equation of state of
the quintessence with
$w_Q(t)=-
(\Lambda+\tau_1\dot\Lambda)/(\Lambda+\tau_2\dot\Lambda)$.
The recent observational bounds on $w_Q$ can be found,
for example, in [14].

I thank A. Ach$\rm \acute u$carro, T.W.B. Kibble, I.I.
Kogan,  L.B. Okun, A.K. Rajantie and A.A. Starobinsky
for fruitful discussions. This work was supported by
ESF COSLAB Programme and by the Russian Foundations for
Fundamental Research.

\end{document}